\documentclass[12pt,twocolum]{elsarticle}
\usepackage{amsmath}
\usepackage{amssymb}
\usepackage{graphicx}
\usepackage{bm}
\journal{Physica B}
\begin{document}
\begin{frontmatter}
\title{Superconductivity on the density wave background with soliton-wall
structure}
\author{P. D. Grigoriev}
\address{L. D. Landau Institute for Theoretical Physics, Chernogolovka, 142432,
Russia; email: grigorev@itp.ac.ru.}
\date{\today }

\begin{abstract}
Superconductivity (SC) may microscopically coexist with density wave (DW)
when the nesting of the Fermi surface (FS) is not perfect. There are, at
least, two possible microscopic structures of a DW state with quasi-particle
states remaining on the Fermi level and leading to the Cooper instability:
(i) the soliton-wall phase and (ii) the small ungapped Fermi-surface
pockets. The dispersion of such quasi-particle states strongly differs from
that without DW, and so do the properties of SC on the DW background. The
upper critical field $H_{c2}$ in such a SC state strongly increases as the
system approaches the critical pressure, where superconductivity first
appears. $H_{c2}$ may considerably exceed its typical value without DW and
has unusual upward curvature as function of temperature. The results
obtained explain the experimental observations in layered organic superconductors
(TMTSF)$_{2}$PF$_{6}$ and $\alpha $-(BEDT-TTF)$_{2}$KHg(SCN)$_{4}$.
\end{abstract}


\end{frontmatter}
\section{Introduction}

\begin{figure}[tb]
\label{PhDia1} \includegraphics[width=0.49\textwidth]{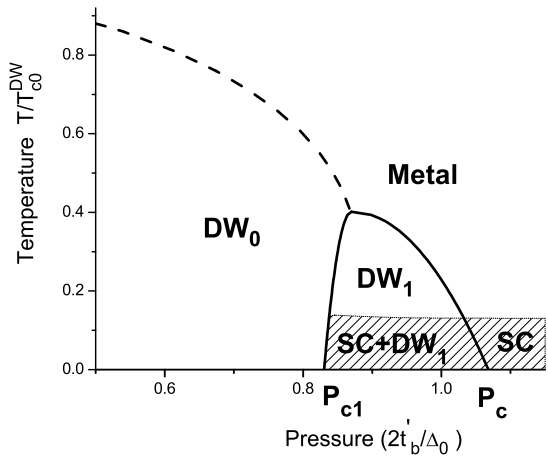}
\caption{The schematic picture of the phase diagram in (TMTSF)$_{2}$PF$_{6}$%
, where superconductivity coexists with DW in some pressure interval above $%
P_{c1}$ but below $P_{c}$. DW$_{0}$ stands for the uniform fully gapped DW.
DW$_{1}$ denotes the DW state when the imperfect nesting term $%
2t_{b}^{\prime }>\Delta _{0}$, so that the ungapped FS pockets or nonuniform
DW structure appear. }
\end{figure}

The interplay of superconductivity (SC) and density-wave (DW) states appears
in many compounds and is extensively investigated for several decades (see,
e.g., reviews in Refs. \cite{Solyom},\cite{Review1}). The DW
removes electrons from the Fermi level, and, usually, precludes
superconductivity\cite{Levin} or strongly reduces the SC transition
temperature\cite{Bilbro}. However, in several compounds [e.g., in layered
organic superconductors (TMTSF)$_{2}$PF$_{6}$ and $\alpha $-(BEDT-TTF)$_{2}$%
KHg(SCN)$_{4}$,\cite{Vuletic,CDWSC} the SC transition temperature $T_{c}^{SC}$ on the DW
background is very close to (or even exceeds) $T_{c}^{SC}$ without DW. In
both these compounds SC coexists with DW in some pressure interval $%
P_{c1}<P<P_{c}$: superconductivity first appears at $P=P_{c1}$, and at $P_{c}
$ the DW phase undergoes a phase transition into the metallic state (see
Fig. 7 in Ref. \cite{Vuletic} and the schematic phase diagram in
Fig. \ref{PhDia1}). In this region SC state has many unusual properties,
such as the divergence of the upper critical field $H_{c2}$ at $P\rightarrow
P_{c1}$.\cite{Hc2Pressure,CDWSC} To explain this property, the scenario of
macroscopic spatial separation of SC and DW states was proposed, where the
size $d_{s}$ of SC domains depends on magnetic field and in strong magnetic
field becomes much smaller than the SC coherence length $\xi _{SC}$.\cite%
{Hc2Pressure} When the width $d_{s}$ of a type II superconducting slab
becomes smaller than $\xi _{SC}$, the upper critical field $H_{c2}$ is
increased compared to $H_{c2}^{0}$ in a bulk superconductor by a factor (see
Eq. 12.4 of Ref. \cite{Ketterson})%
\begin{equation}
H_{c2}/H_{c2}^{0}\approx \sqrt{12}\xi _{SC}/d_{s}.  \label{Hslab}
\end{equation}%
For (TMTSF)$_{2}$PF$_{6}$ and for many other DW superconductors this
scenario implies the size of the SC domains to be of the order of the DW
coherence length $\xi _{DW}$, which means a strong influence of the DW order
parameter on the SC properties and a microscopic coexistence of these two
states. Such a strong spatial modulation of the DW order parameter costs
energy $W$ of the order of the DW energy gap $\Delta _{0}$, which is much
larger than the energy gain due to the surviving of SC state. In this case,
the soliton wall scenario\cite{BrazKirovaReview,SuReview,BGL,GL,GG,GGPRB2007}
is more favorable, where the energy loss $W$ due to nonuniform DW order
parameter is compensated by the large kinetic energy of soliton-band
quasiparticles.\cite{BGL,GG} SC appearing in the soliton wall phase of
spin-density wave (SDW) has the triplet SC pairing\cite{GGPRB2007} in
agreement with experiments in (TMTSF)$_{2}$PF$_{6}$.\cite{LeeTripletMany}
Another microscopic structure was proposed\cite{GGPRB2007} and investigated%
\cite{GrigorievPRB2008} recently, where the small ungapped Fermi-surface
pockets in the DW state appear due to the imperfect nesting of Fermi surface
(FS) and lead to the SC instability. In that scenario the divergence of
upper critical field $H_{c2}$ at $P\rightarrow P_{c1}$ is due to the
decrease of the mean square velocity on the Fermi surface when the size of
the new FS pockets decreases.\cite{GrigorievPRB2008} In the present paper,
we continue to study the SC properties on the DW background. We calculate $%
H_{c2}$ in the soliton-wall scenario and show how the divergence of $H_{c2}$
at $P\rightarrow P_{c1}$ appears on the soliton-phase background. Within
this scenario we also explain the upward curvature of the temperature
dependence of $H_{c2z}$ along the $z$-axis perpendicular to the conducting
layers. The possibility to explain the same feature within the open-pocket
scenario and the limitations of the weak-coupling theory to describe this
effect are briefly discussed. We also explain the observed hysteresis in the
pressure dependence of SC transition temperature $T_{c}^{SC}\left( P\right) $
in both above scenarios.

\section{The model and the DW state without superconductivity}

Below we consider the same model, as in Ref. \cite{GGPRB2007}. The quasi-1D
free electron dispersion without magnetic field has the form
\begin{equation}
\varepsilon (\boldsymbol{k})=\pm v_{F}(k_{x}\mp k_{F})+t_{\perp }(\mathbf{k}%
_{\perp }),  \label{1}
\end{equation}%
where the interchain dispersion $t_{\perp }({\boldsymbol{k}}_{\perp })$ is
much weaker than the in-plane Fermi energy $v_{F}k_{F}$ and given by the
tight-binding model with few leading terms:
\begin{equation}
t_{\perp }({\boldsymbol{k}}_{\perp })=2t_{b}\cos (k_{y}b)+2t_{b}^{\prime
}\cos (2k_{y}b).  \label{dispersion}
\end{equation}%
Here $b$ is the lattice constants in the $y$-direction. The dispersion along
the $z$-axis is considerably weaker than along the $y$-direction and is
omitted. The FS consists of two warped sheets and possesses an approximate
nesting property, $\varepsilon (\boldsymbol{k})\approx -\varepsilon (%
\boldsymbol{k}\mathbf{-}\boldsymbol{Q}_{N})$, with $\boldsymbol{Q}_{N}$\
being the nesting vector. The nesting property leads to the formation of DW
at low temperature and is only spoiled by the second term $t_{b}^{\prime
}(k_{y}\mathbf{)}$ in Eq. (\ref{dispersion}), which, therefore, is called
the "antinesting" term. Increase of the latter with applied pressure leads
to the transition in the DW$_{1}$ state at $P>P_{c1}$, where the
quasi-particle states on the Fermi level first appear and lead to the SC
instability. In the pressure interval $P_{c1}<P<P_{c}$ the new state
develops, where the DW coexists with superconductivity at rather low
temperature $T<T_{c}^{SC}$, while at higher temperature $T_{SC}<T<T_{DW}$
the DW state coexists with the metallic phase. This coexistence takes place
via the formation of small ungapped pockets\cite{GrigorievPRB2008} or via
the soliton phase\cite{GL,GG,GGPRB2007}. We take the DW transition
temperature to be much greater than the SC transition temperature, $%
T_{c}^{DW}\gg T_{c}^{SC}$, which corresponds to most DW superconductors. For
example, in (TMTSF)$_{2}$PF$_{6}$ $T_{c}^{SDW}\approx 8.5K\gg
T_{c}^{SC}\approx 1.1K$, and in $\alpha $-(BEDT-TTF)$_{2}$KHg(SCN)$_{4}$, $%
T_{c}^{CDW}\approx 8K\gg T_{c}^{SC}\approx 0.1K$. Therefore, we first study
the structure of the DW state in the pressure interval $P_{c1}<P<P_{c}$, and
then consider the superconductivity on this background.

In the soliton phase, which presumably establishes in the pressure interval $%
P_{c1}<P<P_{c}$, the DW order parameter depends on the coordinate along the
conducting chains: $\Delta \left( x\right) \approx \Delta _{0}sn\left( x/\xi
_{DW}\right) $, where ${\text{sn}}(y)$ is the elliptic sinus function. As a
result an array of soliton walls of width $\xi _{DW}$ get formed, where the
DW order parameter changes sign. Each soliton wall contributes one
electron-like quasiparticle per conducting chain on the Fermi level, which
leads to the formation of the new conducting soliton band. This soliton 
band appears in the middle of the DW energy gap and has the dispersion
\begin{equation}
E\left( \mathbf{k}\right) =E\left( k_{x}\right) +\varepsilon _{+}(k_{y}),
\label{SolDisp}
\end{equation}%
where the interchain part of the dispersion is given by the antinesting term
in the dispersion (\ref{dispersion}): $\varepsilon _{+}(\mathbf{k}_{\perp })=%
\left[ t_{\perp }({\mathbf{k}}_{\perp })+t_{\perp }({\mathbf{k}_{\perp }-%
\mathbf{Q}_{\perp }})\right] /2.$ The dispersion $E\left( k_{x}\right) $
along the conducting chains was found in Ref. \cite{BGK} (see Fig. 1 in Ref.
\cite{BGK}), and for qualitative analysis it can be approximated by%
\begin{equation}
E\left( k_{x}\right) \approx E_{-}\sin \left[ \pi \left( \left\vert
k_{x}\right\vert -k_{F}\right) /2\kappa _{0}\right] .  \label{EsAppr}
\end{equation}%
The soliton band width $E_{-}$ and boundary $\kappa _{0}$ in the momentum
space are determined by the linear concentration $n_{s}$ of the soliton walls%
\cite{BGK}: $\kappa _{0}=\pi n_{s}/2$, $\hbar v_{F}n_{s}=E_{+}/K\left( \sqrt{%
1-E_{-}^{2}/E_{+}^{2}}\right) $, where $K(r)$ is the complete elliptic
integral of the 1st kind. At $n_{s}\rightarrow 0$ $E_{+}\approx \Delta _{0}$,
 and $E_{-}\approx 4\Delta _{0}\exp \left( \ -\Delta _{0}/\hbar
v_{F}n_{s}\right) $. Note, that the soliton phase can be energetically most
favorable DW state at imperfect nesting.\cite{GG} Compared to the uniformly gapped
DW state, the soliton phase gains the kinetic energy of quasiparticles in
the half-filled soliton phase due to the term $\varepsilon _{+}(k_{y})$ in (%
\ref{SolDisp}), which compensates the energy loss of the non-uniform DW
structure.

The soliton wall concentration $n_{s}$ depends on external pressure $P$
via the dependence on electron dispersion. $n_{s}\left( P\right) $ can
be obtained by minimization of the total energy of the soliton phase
(see Eqs. (33)-(35) of Ref. \cite{BGL} or Eqs. (9)-(12) of Ref. \cite%
{GG}). The result substantially depends on the harmonic content of electron
dispersion $t_{\perp }\left( \mathbf{k}_{\perp }\right) $\cite{GG}. For
typical dispersion, at $P\rightarrow P_{c1}$ the soliton band width $%
E_{-}/\Delta _{0}\rightarrow 0$, which corresponds to a phase transition
from the uniform DW to the soliton phase. In the vicinity of $P=P_{c1}$,
rough estimates give\cite{GP}
\begin{equation}
E_{-}\sim \sqrt{\Delta _{0}\delta }\equiv \sqrt{\Delta _{0}\left(
2t_{b}^{\prime }-\Delta _{0}\right) }\propto \sqrt{P-P_{c1}}.  \label{Emnd}
\end{equation}

\section{SC properties on the soliton-phase DW background}

In the Ginzburg-Landau functional for SC state the coefficient tensor $%
1/m_{ij}$ before the gradient term can be expressed via the product $%
\left\langle v_{i}v_{j}\right\rangle $ of electron velocities $v_{i}$
averaged over the Fermi surface\cite{Melik} [see Eqs. (48)-(50) of Ref. \cite%
{GrigorievPRB2008}. Using this result and the simplified dispersion 
(\ref{SolDisp}),(\ref{EsAppr}) in the soliton band one easily obtains
\begin{eqnarray}
\frac{1}{m_{yy}} &=&\frac{14\zeta \left( 3\right) \left( t_{b}^{\prime
}b\right) ^{2}}{3\pi ^{2}T_{c}^{SC}\hbar ^{2}},  \label{msol} \\
\frac{1}{m_{xx}} &=&\frac{7\zeta \left( 3\right) E_{-}^{2}\left( \pi
/2\kappa _{0}\right) ^{2}}{24\pi ^{2}T_{c}^{SC}\hbar ^{2}}=\frac{7\zeta
\left( 3\right) E_{-}^{2}}{24\pi ^{2}T_{c}^{SC}\hbar ^{2}n_{s}^{2}},  \notag
\end{eqnarray}%
and the upper critical field $H_{c2}^{z}=\left( c/e\hbar \right) \left(
T_{c}^{SC}-T\right) \sqrt{m_{xx}m_{yy}}$ along the z-axis is%
\begin{equation}
H_{c2}^{z}=C_{1s}\frac{T_{c}^{SC}\left( n_{s}\xi _{SDW}\right) }{E_{-}}\frac{%
c\left( T_{c}^{SC}-T\right) }{ebv_{F}},  \label{Hc2z2}
\end{equation}%
where $\xi _{DW}=\hbar v_{F}/\pi \Delta _{0}$, $n_{s}\xi _{DW}\approx 1/\pi
\ln \left( 4\sqrt{2}\Delta _{0}/E_{-}\right) $, and the constant $%
C_{1s}=12\pi ^{3}/7\zeta \left( 3\right) \approx 44.2.$ Eq. (\ref{Hc2z2}) is
similar to Eq. (55) of Ref. \cite{GrigorievPRB2008} for the second scenario,
where the open pocket size $\delta =2t_{b}^{\prime }-\Delta _{0}$ is
replaced by the soliton band width $E_{-}$. At $P\rightarrow P_{c1}~$both $%
\delta \rightarrow 0$ and $E_{-}\rightarrow 0$, and in both scenarios the
functions $\delta \left( P\right) $ and $E_{-}\left( P\right) $ depend on
the electron dispersion $t_{\perp }\left( \mathbf{k}_{\perp }\right) $.
Substituting (\ref{Emnd}) into (\ref{Hc2z2}), one obtains that in the
soliton-wall scenario the slope $dH_{c2}^{z}/dT\propto 1/\sqrt{P-P_{c1}}$
and the upper critical field $H_{c2}$ diverge at $P\rightarrow P_{c1}$
similar to the scenario of the ungapped FS pockets\cite{GrigorievPRB2008}. 
The physical reason for this divergence is the strong change of quasi-particle 
dispersion on the Fermi level when $P\to P_{c1}$ and the soliton band shrinks.
This shrinking leads to the decrease in the mean square electron velocity on 
the Fermi level (or diffusion coefficient), which increases the upper critical 
field $H_{c2}$. 

For (TMTSF)$_{2}$PF$_{6}$ the substitution of $E_{-}\sim \sqrt{\Delta
_{0}\delta }$ to (\ref{Hc2z2}) gives the slope
\begin{equation}
\frac{dH_{c2}^{z}}{dT}\approx \frac{7.8}{\ln \left( 4\sqrt{2\Delta
_{0}/\delta }\right) }\frac{T_{c}^{SC}}{\sqrt{\Delta _{0}\delta }}\left[
\frac{Tesla}{^{\circ }K\,}\right] .  \label{Slope1}
\end{equation}%
and the maximum slope $dH_{c2}^{z}/dT\approx $ $0.6\,\left[ Tesla/^{\circ }K%
\right] $ in a reasonable agreement with the experiment (see Fig. 2 in Ref. 
\cite{Hc2Pressure}). Far from $P=P_{c1}$, $E_{-}\approx \Delta _{0}$,
and\ Eq. (\ref{Hc2z2}) gives $dH_{c2}^{z}/dT\approx $ $0.25\,\left[
Tesla/^{\circ }K\right] $, which is again in a good agreement with
experimental data in Fig. 2 of Ref. \cite{Hc2Pressure}. For $\alpha $%
-(BEDT-TTF)$_{2}$KHg(SCN)$_{4}$, the substitution of $E_{-}\approx \Delta
_{0}$ to (\ref{Hc2z2}), using BCS relation $\Delta _{0}=1.76T_{c}^{CDW}$,
gives $H_{c2}^{z}\approx $ $7.7\,$mTesla$\,\cdot \left(
1-T/T_{c}^{SC}\right) $, or $dH_{c2}^{z}/dT\approx $ $77\,\left[
mTesla/^{\circ }K\right] $ in a qualitative agreement with experiment\cite%
{CDWSC}.

The upward curvature of $H_{c2}^{z}(T)$ in the scenario of soliton walls is
similar to $H_{c2}(T)$ in layered superconductors when magnetic field is
parallel to the conducting layers. The latter case was considered in a
number of theoretical papers (see, e.g., \cite{Lawrence},\cite{Klemm},\cite%
{Deutscher}, or \S 16 of \cite{Ketterson}). The 1D network of soliton walls
can be considered as a 1D Josephson lattice, where the conducting layers of
thickness $\xi _{DW}$ are separated by the insulating layers by thickness $%
s=1/n_{s}$. At low field this Josephson lattice behaves like a 3D
superconductor with critical field $H_{c2}^{i}=e_{ijk}\Phi _{0}/2\pi \xi
_{j}\left( T\right) \xi _{k}\left( T\right) ,$ where $\Phi _{0}$ is magnetic
flux quantum and $\xi _{i}\left( T\right) $ is the SC correlation length in $%
i$-direction.  In the vicinity of critical temperature $\xi _{i}\propto
(T_{cSC}-T)^{-1/2}$. At high field $H_{c2}$ tends to its
value (\ref{Hslab}) in the 2D SC slab of thickness $d_{s}\sim \xi _{DW}\ll
\xi _{SC}$.\cite{Deutscher} For (TMTSF)$_{2}$PF$_{6}$, $\xi _{DW}/\xi
_{SC}\approx 1/10$, and according to Eq. (\ref{Hslab}), the upper critical
field $H_{c2}^{z}$ on the SDW background can be enhanced by a factor $%
H_{c2}/H_{c2}^{0}\lesssim 30$ as compared to $H_{c2}^{z0}$ in the bulk
superconductor, i.e. superconductor without the soliton structure.

The crossover from the 3D to 2D behavior of $H_{c2}^{z}$ occurs when the
soliton bandwidth becomes small: $E_{-}<\hbar ^{2}/m_{y}\xi _{SC}^{2}$,
where $\xi _{SC}\approx \hbar v_{y\max }/\pi \Delta _{SC}\left( T\right) $
is the temperature dependent correlation length within the conducting layer,
and $m_{y}\approx t_{b}^{\prime }/v_{y\max }^{2}$. This gives the crossover
value $E_{-}\sim \lbrack \pi \Delta _{SC}\left( T\right) ]^{2}/t_{b}^{\prime
}$. Therefore, the upper critical field increases and behaves as in the
isolated SC slab only at $P\rightarrow P_{c1}$ and only at low temperature,
which means the unusual upward curvature of $H_{c2}^{z}\left( T\right) $ at $%
P\rightarrow P_{c1}$. Note, that the soliton structure is important for the
upward curvature of only z-component $H_{c2}^{z}\left( T\right) $ of
magnetic field in layered organic metals as (TMTSF)$_{2}$PF$_{6}$ or $\alpha
$-(BEDT-TTF)$_{2}$KHg(SCN)$_{4}$, because it creates the layered soliton
wall structure parallel to the magnetic field. If magnetic field lies in the
x-y plane, it is already parallel to the conducting molecular layers, and 
$H_{c2}^{z}(T)$ may have the upward curvature according to the Lawrence-Doniach model
even without soliton walls.

\section{Discussion and summary}

The above study gives a rough estimates of the upper critical field $%
H_{c2}\left( P,T\right) $ in the soliton phase. The divergence and the
unusual upward curvature of $H_{c2}^{z}\left( T\right) $ at $P\rightarrow
P_{c1}$, observed in (TMTSF)$_{2}$PF$_{6}$ and $\alpha $-(BEDT-TTF)$_{2}$%
KHg(SCN)$_{4}$, are explained. A quantitative study of $H_{c2}\left(
P,T\right) $ requires a more accurate model for the coexisting SC and DW
states, which includes the influence of SC on the DW state. The critical
fluctuations near the DW-metal phase transition also strongly influence the
SC state via the renormalization of the e-e interaction.\cite{Solyom,DupuisReview,Kuroki} 
However, the proposed rough model gives a
reasonable quantitative agreement with experiment on the slope of $%
H_{c2}^{z}\left( T\right) $ at $T\rightarrow T_{c}^{SC}$. 

One more unusual feature observed in these compounds is the hysteresis in
the pressure dependence of the SC transition temperature $T_{c}^{SC}\left(
P\right) $. In the soliton scenario this hysteresis may come from the motion
of soliton walls in order to achieve the optimal soliton-wall concentration $%
n_{s}\left( P\right) $ as pressure $P$ changes. The hysteresis in $%
T_{c}^{SC}\left( P\right) $ can also be explained in the open-pocket
scenario if the DW wave vector shifts at $P>P_{c1}$ with change in pressure
to find the optimal DW wave vector at imperfect nesting. Since the DW wave
vector is pinned by crystal imperfections, this shift of the DW wave vector
is hysteretic, which leads to the hysteresis in $\delta $ and in the SC
transition temperature $T_{SC}$.

An accurate description of the SC properties in both scenarios must go
beyond the weak-coupling theory, because the new Fermi energy (i.e., $\delta $ in
the new FS pockets or the bandwidth $E_{-}$ in the soliton phase) is
comparable to the SC energy gap as $P\rightarrow P_{c1}$ and is much smaller
than the Debye energy. This fact, for example, may be responsible for the
upward curvature of $H_{c2}^{z}\left( T\right) $ in the scenario of open
pockets, where
\begin{equation}
H_{c2}^{z}\propto 1/\sqrt{\left\langle v_{x}^{2}\right\rangle \left\langle
v_{y}^{2}\right\rangle }\propto 1/\delta .  \label{Hc2delta}
\end{equation}
When $\delta <T_{c}^{SC}$, the small value of $\delta $ is smeared out by
high temperature. Qualitatively, this smearing may be included via the
replacement $\delta \rightarrow \sqrt{\delta _{0}^{2}+\alpha T^{2}}$, $%
\alpha \sim 1$. At $P\rightarrow P_{c1}$ substituting this and $\delta
_{0}\propto P-P_{c1}$ to (\ref{Hc2delta}), one obtains
\begin{equation}
H_{c2}^{z}\propto 1/\sqrt{\beta \left( P-P_{c1}\right) ^{2}+\alpha T^{2}},
\label{Hc2UpPocket}
\end{equation}%
which gives the upward curvature of $H_{c2}^{z}\left( T\right) $ and
describes well the experimental data in Ref. \cite{Hc2Pressure} with two
fitting parameters $\alpha $ and $\beta $.

The proposed approximate model of the SC state on the soliton-wall
background of the DW at $P>P_{c1}$ qualitatively explains all unusual SC
properties on the DW background observed in the organic metals (TMTSF)$_{2}$%
PF$_{6}$ and $\alpha $-(BEDT-TTF)$_{2}$KHg(SCN)$_{4}$, which gives a strong
argument in favour of the microscopic coexistence of superconductivity and
density-wave state in these compounds.

The work was supported by RFBR No 06-02-16551, No 06-02-16223, by
MK-4105.2007.2 and the research program of JSC "Advanced Energetic
Technologies".

\end{document}